\documentclass[12pt]{article}
\usepackage{graphicx,epsfig}
\usepackage{color}
\usepackage{epsfig,float,amsmath,amsfonts,amssymb,graphicx,bm,bbm,array,dcolumn}

\def\beq{\begin{equation}}
\def\eeq{\end{equation}}
\def\bea{\begin{eqnarray}}
\def\eea{\end{eqnarray}}
\def\bmat{\begin{pmatrix}}
\def\emat{\end{pmatrix}}

\newcommand{ \slashchar }[1]{\setbox0=\hbox{$#1$}   
   \dimen0=\wd0                                     
   \setbox1=\hbox{/} \dimen1=\wd1                   
   \ifdim\dimen0>\dimen1                            
      \rlap{\hbox to \dimen0{\hfil/\hfil}}          
      #1                                            
   \else                                            
      \rlap{\hbox to \dimen1{\hfil$#1$\hfil}}       
      /                                             
   \fi}                                             %

\def\to{\rightarrow}

\begin{document}

\title{
LHC Signatures of a Vector-like $b'$
}

\author{S.~Gopalakrishna$^{a}$\thanks{shri@imsc.res.in}, T.~Mandal$^a$\thanks{tanumoy@imsc.res.in}, S.~Mitra$^b$\thanks{smitra@iopb.res.in}, R.~Tibrewala$^a$\thanks{rtibs@imsc.res.in} \\
$^a$~\small{Institute of Mathematical Sciences (IMSc), Chennai 600113, India.}\\
$^b$~\small{Institute of Physics (IOP), Bhubaneswar 751005, India.}}

\date{}

\maketitle

\begin{abstract}
Many beyond the standard model extensions predict the existence of heavy vector-like fermions.  
We study the LHC signatures of one such heavy vector-like fermion, called $b'$, with electromagnetic charge $-1/3$
like the SM $b$-quark,
but which could generically have different $SU(2)_L$ and $U(1)_Y$ quantum numbers.
Our emphasis will be on the phenomenology due to $b \leftrightarrow b'$ mass-mixing, 
present after electroweak symmetry breaking.
We focus on aspects which distinguish a vector-like $b'$ from a chiral $b'$ and include tree-level decays of the $b'$ 
into $t W$, $b Z$ and $b h$ final states.
While our analysis is largely model-independent, we take as a motivating example warped-space models    
in which a vector-like $b'$ appears as the custodial partner of the top-quark.
\end{abstract}

\section{Introduction}
\label{Intro.SEC}
The standard model (SM) of particle physics suffers from the gauge hierarchy and flavor hierarchy problems and
many extensions have been proposed to solve these problems. 
These theories beyond the standard model (BSM) predict extra particles that are being searched for at the 
CERN Large Hadron Collider (LHC). 
Among them are extra heavy fermions which could either be vector-like or chiral.
The purpose of this study is to analyze the LHC signatures of a vector-like fermion which has
electromagnetic charge $-1/3$, but depending on the model, can have various $SU(2)_L$ and $U(1)_Y$ quantum numbers.
We refer to such a state as a $b'$. 
By vector-like we mean that in the theory are present both a state in a representation of the gauge group and its 
conjugate representation, while a chiral state is one for which its conjugate representation is not present. 
A vector-like fermion can have a bare mass consistent with gauge invariance, in contrast to a chiral fermion that 
obtains its mass due to the (spontaneous) breaking of the gauge symmetry.

The vector-like nature of the $b'$ can ascribe certain unique features to it which distinguishes it from a 
chiral $b'$ that obtains its mass due to the SM Higgs vacuum expectation value (VEV).  
In the chiral case, the dominant decay mode of the $b'$ is likely to be into $t W$, unless the
$t' W$ is kinematically accessible (where the $t'$ is a charge $2/3$ heavy fermion), 
induced by the charged current interaction.  
Also, in the chiral case, diagonalizing the mass matrix automatically diagonalizes the Higgs interactions, 
due to which a $b' b h$ coupling is absent at tree-level, while for a vector-like theory, 
diagonalizing the mass matrix does not render the Higgs interactions diagonal due to the presence 
of the vector-like mass term that is independent of the Higgs VEV.  
Thus, a vector-like $b'$ can have a tree-level $b' \to b h$ decay, which is not present at tree-level for chiral $b'$ 
that gets its mass solely from the Higgs VEV.
Furthermore, if the $SU(2)_L$ and $U(1)_Y$ quantum numbers of the $b'$ are not the same as the $b$, 
generically, after electroweak symmetry breaking (EWSB), an off-diagonal $b' b Z$ vertex is generated
which allows the tree-level $b' \to b Z$ decay. Depending on the model, this branching ratio (BR)
can be much bigger compared to a chiral $b'$ that has identical $SU(2)_L$ and $U(1)$ quantum numbers as the $b$ 
(for example, a fourth generation extension of the SM), for which this decay occurs only at loop level.  
At the LHC, largish $b' \to b Z$ and $b' \to b h$ BRs will reveal these aspects of the $b'$.  
In this work, therefore, we pay particular attention to the $b Z$ and $b h$ decay modes along with the $t W$ decay mode, 
obtain $b'$ cross-sections for the significant
production channels at the LHC, and analyze the reach for a vector-like $b'$.  
We present our results quite model-independently, but motivate our analysis in the context of warped-space models. 

Chiral heavy fermions have been studied well in the literature, particularly in the context of fourth generation models. Here we will study
a vector-like $b'$ model-independently, but keeping in mind, as an example, warped extradimensional theories~\cite{rs1}
that have been proposed to solve the gauge-hierarchy problem.
Due to the AdS/CFT correspondence conjecture~\cite{Maldacena:1997re}, these are dual to 4-dimensional strongly-coupled theories.  
In variants of the original warped extradimensional proposal, the custodial partners of the top-quark (including the $b'$) can be significantly 
lighter~\cite{Choi:2002ps,Agashe:2003zs,Agashe:2004bm,Agashe:2004cp} than all the other Kaluza-Klein (KK) particles, 
making its observability at the LHC promising. 
Various studies have considered the LHC signatures of such TeV scale vector-like fermions. 
Ref.~\cite{Dennis:2007tv} considers the LHC signatures of a vector-like $b'$ by looking at 4-$W$ events, along with signatures of 
a charge~$5/3$ fermion, 
Ref.~\cite{Contino:2008hi} considers the single and pair-production of the charge $5/3$ custodial partner of the SM left-handed quark doublet
exploiting same-sign dileptons to beat SM background, and the same-sign signal is also considered in Ref.~\cite{Mrazek:2009yu}. 
Ref.~\cite{AguilarSaavedra:2009es} studies pair-production followed by decays into single and multi-lepton channels,
and the pair-production of the KK top is explored in Ref.~\cite{Carena:2007tn}. 
Signals due to mixing with light quarks and constraints have been analyzed in Ref.~\cite{Atre:2008iu}.  
In Ref.~\cite{Brooijmans:2010tn} an exhaustive list of the single $b'$ production processes at the LHC was given for the first time 
and new dominant processes were pointed out. This work draws heavily from the investigations there, 
which are being studied further~\cite{Gopalakrishna:2009xxx}. 
For this model, the partial decay-widths are worked out in Ref.~\cite{Ghosh:2011zz}.
On the experimental front, the Tevatron (CDF) bound is presented in Ref.~\cite{Aaltonen:2007je} and is also discussed in Ref.~\cite{Hung:2007ak}. 
Recent LHC (CMS) bounds from the $b'\to t W$ decay mode is presented in Ref.~\cite{Chatrchyan:2011em}. 
Our emphasis here will be to include $b'$ single and pair-production, and, $b' \to b Z$ and $b' \to b h$ decay modes in addition to $b' \to tW$, 
and keeping it model-independent.
Single-production depends more directly on the electroweak quantum numbers of the $b'$, while $b'$ pair-production is dominated by its 
coupling to the gluon (which is given by the $SU(3)_C$ gauge coupling $g_S$) and thus hides its electroweak nature.
For this reason, in addition to pair-production, we will consider single-production also in our work.  

The paper is organized as follows:
In Sec.~\ref{bpMixCoupl.SEC} we present the effective Lagrangian in a model-independent way showing
the coupling of the $b'$ to SM fields, and identify the relevant parameters for our work.
In Sec.~\ref{bpDecay.SEC} we derive expressions for the $b'$ partial decay widths. 
In Sec.~\ref{LHCSign.SEC} we explore the $b'\bar{b'}$ pair production and $b'Z$, $b' h$ single production, particularly focussing on $bZ$ and $bh$
decays going into the semileptonic and dileptonic final-states, compute signal and background cross-sections with appropriate cuts, 
and compute the luminosity required at the 14~TeV LHC for $5\sigma$ significance with at least 10~events.
For the semileptonic mode, we include the dominant QCD background, in addition to the irreducible electroweak background. 
We present these results model-independently by varying the relevant couplings and the $b'$ mass.
We list a few other $b'$ single production processes very briefly and mention the reasons why we do not consider them in detail.  
In Sec.~\ref{CONCL.SEC} we offer our conclusions.

\section{$b'$ mass-mixing and couplings}
\label{bpMixCoupl.SEC}
We consider an extension of the SM with a heavy (TeV scale) vector-like $b'$.
Generically, after EWSB, the SM $b$ mixes with the $b'$ due to
off-diagonal terms in the mass matrix. 
After taking into account this mixing, we can go from the $(b,b')$ basis to the $(b_1,b_2)$ mass-basis
and write the Lagrangian model-independently in the mass-basis as
\bea
{\cal L}_{4D} \supset 
&-&\frac{e}{3} \bar{b_1} \gamma^\mu b_1 A_\mu 
-\frac{e}{3} \bar{b}_{2} \gamma^\mu b_{2} A_\mu
+ g_s \bar{b_1} \gamma^\mu T^\alpha b_1 g^\alpha_\mu  
+ g_s \bar{b}_{2}  \gamma^\mu T^\alpha b_{2} g^\alpha_\mu \nonumber \\
&-& \left( \kappa^L_{b t W} \bar{t_L} \gamma^\mu {b_1}_L W_\mu^+ 
+ \kappa^L_{b_2 t W} \bar{t}_{1L}  \gamma^\mu {b_2}_L W_\mu^+ + {\rm h.c.} \right) \nonumber \\
&+& \kappa_{b b Z}^L \bar{b_1}_L \gamma^\mu {b_1}_L Z_\mu
+ \kappa_{b_2 b_2 Z}^L \bar{b}_{2L} \gamma^\mu b_{2L} Z_\mu \nonumber \\
&+& \left( \kappa_{b_2 b Z}^L \bar{b}_{1L} \gamma^\mu b_{2L}  Z_\mu + {\rm h.c.} \right)  \nonumber \\
&+& \kappa_{b b Z}^R \bar{b_1}_R \gamma^\mu {b_1}_R Z_\mu 
+ \kappa_{b_2 b_2 Z}^R \bar{b}_{2R}  \gamma^\mu b_{2R}  Z_\mu \ ,
\label{b'uni.EQ}
\eea
and the Higgs interactions as
\bea
{\cal L}_{4D} \supset -\frac{h}{\sqrt{2}} \left[ 
\kappa_{h b_{L} b_{R}} \bar{b}_{1L} {b}_{1R} + 
\kappa_{h b_{2L} b_{2R}} \bar{b}_{2L} {b}_{2R} \right. \nonumber \\
\left. + \kappa_{h b_{L} b_{2R}} \bar{b}_{1L} {b_2}_R +
\kappa_{h b_{2L} b_{R}} \bar{b}_{2L} {b_1}_R  \right] + {\rm h.c.} \ . 
\label{Lagb'H.EQ}
\eea
We omit a few other possible terms in Eq.~(\ref{b'uni.EQ}) for the following reasons: 
our interest will be in theories in which the mass mixing is between $b_L \leftrightarrow b'_L$
without $b_R$ mixing, which is the reason why we do not introduce 
$\kappa_{b_2 b Z}^{R} \bar{b}_{2R}  \gamma^\mu b_{1R}  Z_\mu + {\rm h.c.} $.
Also, we will assume that the $W_L \leftrightarrow W_R$ mixing is small (where the $W_R$ is the $SU(2)_R$
gauge boson if this symmetry is gauged), and that the $b'$ is a singlet under $SU(2)_L$, 
which is why we do not include a
$ \kappa^R_{b_2 t W} \bar{t}_{1R}  \gamma^\mu {b_2}_R W_\mu^+ + {\rm h.c.}$ term.  
For convenience, in the text, we use $(b, b')$ interchangeably with the mass eigenstates $(b_1,b_2)$, 
but in our numerical work we distinguish them properly. 

Here we consider a single $b'$ for simplicity, but in general, 
there could be more than one $b'$ that mixes with the $b$, 
and our work can be straightforwardly extended to 
models with more than one $b'$.\footnote{Unless the $SU(2)_L \otimes U(1)_Y$ quantum numbers of the $b'$ are the same as that of the $b$, 
mixing to a single $b'$ puts a stringent lower bound on the $b'$ mass ($M_{b'} \gtrsim 3$~TeV) from the 
requirement that the shifts to the $Zb\bar b$ coupling be smaller than the constraint from precision electroweak data. 
For example, in the warped-space model in Ref.~\cite{Brooijmans:2010tn} with the $b$ mixing to a
single $b'$ with {\em different} $SU(2)_L\otimes U(1)_Y$ quantum numbers, we explicitly see that the 
the $Zb\bar b$ coupling gets shifted.  
This can be avoided by either ensuring that the $SU(2)_L \otimes U(1)_Y$ quantum numbers of the $b'$ is the
same as that of the $b$, or, by mixing to more than one $b'$. 
The latter is the case for instance for the warped-space model in Ref.~\cite{Agashe:2006at}. 
}

Here, we present the phenomenology in a model-independent manner, 
and vary the mass of the $b'$ (denoted as $M_{b_2}$) in presenting the phenomenology. 
We stipulate that the underlying BSM model must ensure that $\kappa_{b b Z}^{L,R}$, $\kappa_{b t W}$ and $\kappa_{h b_{L} b_{R}}$,
to a good approximation,
take their respective SM values to be consistent with experimental data.
In the BSM extensions we are interested in, $\kappa_{h b_{2L} b_{R}}$ will be very small and therefore
we set this to zero in our analysis. 
We take the remaining $\kappa$'s, namely, 
$\kappa_{b_2 b Z}^{L}$, $\kappa_{b_2 t W}$, $\kappa_{h b_{L} b_{2R}}$, $\kappa_{b_2 b_2 Z}^{L,R}$, and $\kappa_{h b_{2L} b_{2R}}$ 
as free parameters;
our phenomenology will only depend on the first three of these $\kappa$'s, and the last two are largely irrelevant here. 
This is because although the $b'$ pair production has contributions due to the last two couplings, they are sub-dominant
compared to the gluon exchange channel, and, the last two couplings are not relevant for single production or decay
of the $b'$. 

We analyze the phenomenology in the following sections for the benchmark masses and couplings shown in Table~\ref{MbpKappa.TAB}.
These are the couplings obtained for the warped-space model considered in Ref.~\cite{Brooijmans:2010tn} where the
$\kappa_{ijk}$ are explicitly worked out for the $b' \subset (1,2)$ representation of $SU(2)_L \otimes SU(2)_R$. 
\begin{table}[!h]
\begin{center}
\caption{The benchmark masses and couplings used in this study. These are obtained for the warped model in Ref.~\cite{Brooijmans:2010tn}.
\label{MbpKappa.TAB} }
\begin{tabular}{|c|c|c|c|c|c|c|c|c|c|c|c|c|}
\hline 
$M_{b_2}$ (GeV) & 250 & 500 & 750 & 1000 & 1250 & 1500  \tabularnewline
\hline
$\kappa_{b_2 b Z}^{L}$ & 0.185 & 0.121 & 0.084 & 0.064 & 0.051 & 0.043 \tabularnewline
\hline 
$\kappa_{b_2 t W}$ & 0.322 & 0.161 & 0.107 & 0.080 & 0.064 & 0.054  \tabularnewline
\hline
$\kappa_{h b_{L} b_{2R}}$ & 0.714 & 0.937 & 0.972 & 0.985 & 0.990 & 0.993  \tabularnewline
\hline
\hline
$M_{b_2}$ (GeV) & 1750 & 2000 & 2250 & 2500 & 2750 & 3000 \tabularnewline
\hline
$\kappa_{b_2 b Z}^{L}$ & 0.037 & 0.032 & 0.029 & 0.026 & 0.024 & 0.022 \tabularnewline
\hline 
$\kappa_{b_2 t W}$ & 0.046 & 0.040 & 0.036 & 0.032 & 0.029 & 0.027 \tabularnewline
\hline
$\kappa_{h b_{L} b_{2R}}$ & 0.995 & 0.996 & 0.997 & 0.998 & 0.998 & 0.998 \tabularnewline
\hline
\end{tabular}
\end{center}
\end{table}

\section{$b^\prime$ decay}
\label{bpDecay.SEC}
\begin{figure}[h!]
\begin{center}
\includegraphics[width=0.32\textwidth]{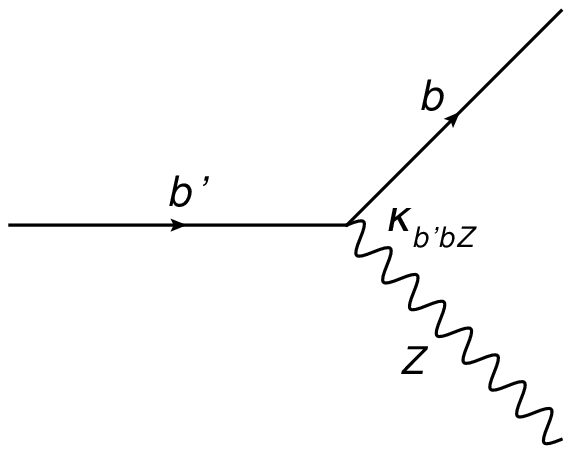}
\includegraphics[width=0.32\textwidth]{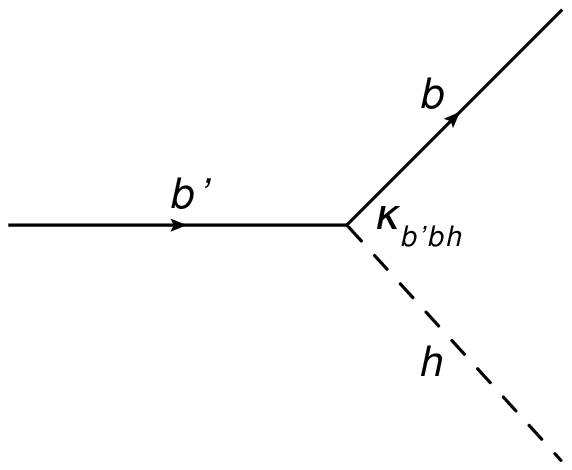}
\includegraphics[width=0.32\textwidth]{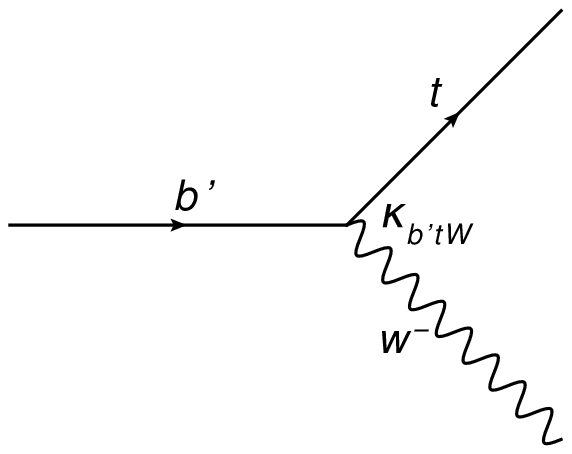}
\caption{The tree-level decay modes of the $b'$.
\label{b2Decays.FIG}}
\end{center}
\end{figure}

The heavy mass eigenstate $b_2$, once produced, decays via the off-diagonal interaction terms 
in Eqs.~(\ref{b'uni.EQ})~and~(\ref{Lagb'H.EQ}). 
Thus, the main decay modes are $b_2\to b_1\, Z$, $b_2 \to b_1\, h$ and $b_2 \rightarrow t\, W$,
and these tree-level decays are shown in Fig.~\ref{b2Decays.FIG}. 
As already mentioned, the presence of the $b_1 h$ decay mode uniquely signals the
vector-like nature of the $b'$. 
The partial-widths into these decay channels are
\begin{eqnarray}
\Gamma_{bZ}\!\!\! &=& \!\!\!
\frac{\left(\kappa^{L}_{{b_2}bZ}\right)^2}{32\pi} M_{b_2}
\cdot \left( \frac{1}{x_Z^2} + 1 - 2 x_{bZ}^2 + x_b^2 - 2 x_Z^2 +
x_{bZ}^2 x_b^2 \right) \times \nonumber \\
&& \left( 1 + x_Z^4 + x_b^4 - 2 x_Z^2 -2 x_b^2 - 2 x_Z^2 x_b^2
\right)^{1/2} \ , \\
\Gamma_{tW} \!\!\! &=& \!\!\!
\frac{\left(\kappa^{L}_{{b_2}tW}\right)^2}{32\pi} M_{b_2}
\cdot \left( \frac{1}{x_W^2} + 1 - 2 x_{tW}^2 + x_t^2 - 2 x_W^2 +
x_{tW}^2 x_t^2 \right) \times \nonumber \\
&&  \left( 1 + x_W^4 + x_t^4 - 2 x_W^2 -2 x_t^2 - 2 x_W^2 x_t^2
\right)^{1/2} \ ,
\label{GamqV.EQ} \\
\Gamma_{bh} \!\!\! &=& \!\!\! \frac{M_{b_2}}{64\pi}
\left[
\left(\kappa_{hb_{2L}b_R}^2 + \kappa_{hb_{L}b_{2R}}^2\right) \left(1 -
x_h^2 - \frac{3}{4}x_b + x_b^2\right)
+ \frac{5}{2}\kappa_{hb_{2L}b_R}\kappa_{hb_{L}b_{2R}} x_b
\right]\nonumber\\
&& \times\left(1+x_h^4 + x_b^4 -2x_h^2 - 2x_b^2 -2x_h^2x_b^2\right)^{1/2},
\label{Gambh.EQ}
\eea
where $x_i = m_i/M_{b_2}$, $x_{ij} = m_i/m_j$, and, 
$\Gamma_{bZ}, \Gamma_{tW}$ and $\Gamma_{bh}$ denote the
partial widths of the $b'$ to the $bZ$, $tW$ and $bh$ final states
respectively.

Since the $b_2 b h$ coupling in Table~\ref{MbpKappa.TAB} is large, $\Gamma_{b h}$ can be sizable.
The $\Gamma_{b Z}$ dependence on $1/x_Z^2$ due to the longitudinal polarization of the $Z_\mu$ enhances this partial width 
for large $M_{b_2}$ and can make it comparable to $\Gamma_{b h}$.
The same holds also for the $\Gamma_{t W}$. 

We show in Fig.~\ref{b2-GamMI.FIG} the partial widths of the $b'$ to the $bZ$, $tW$ and $bh$ final states,
in a model-independent fashion, in the $\kappa$ -- $M_{b_2}$ plane. 
The blue dots show the relation between the $M_{b_2}$ and $\kappa$ as shown in Table~\ref{MbpKappa.TAB}, 
and the partial-widths in this model can be read-off from the plots. 
\begin{figure}[h!]
\begin{center}
\includegraphics[width=0.32\textwidth]{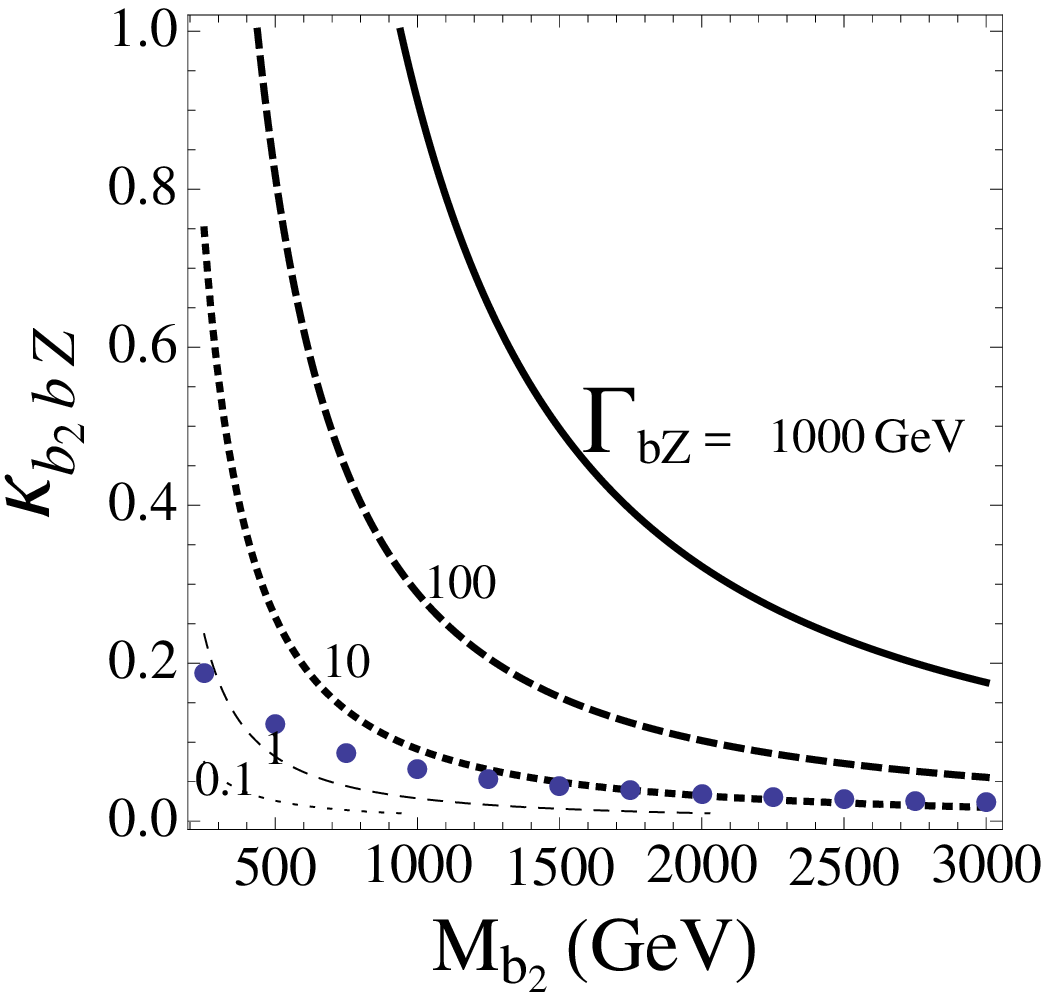}
\includegraphics[width=0.32\textwidth]{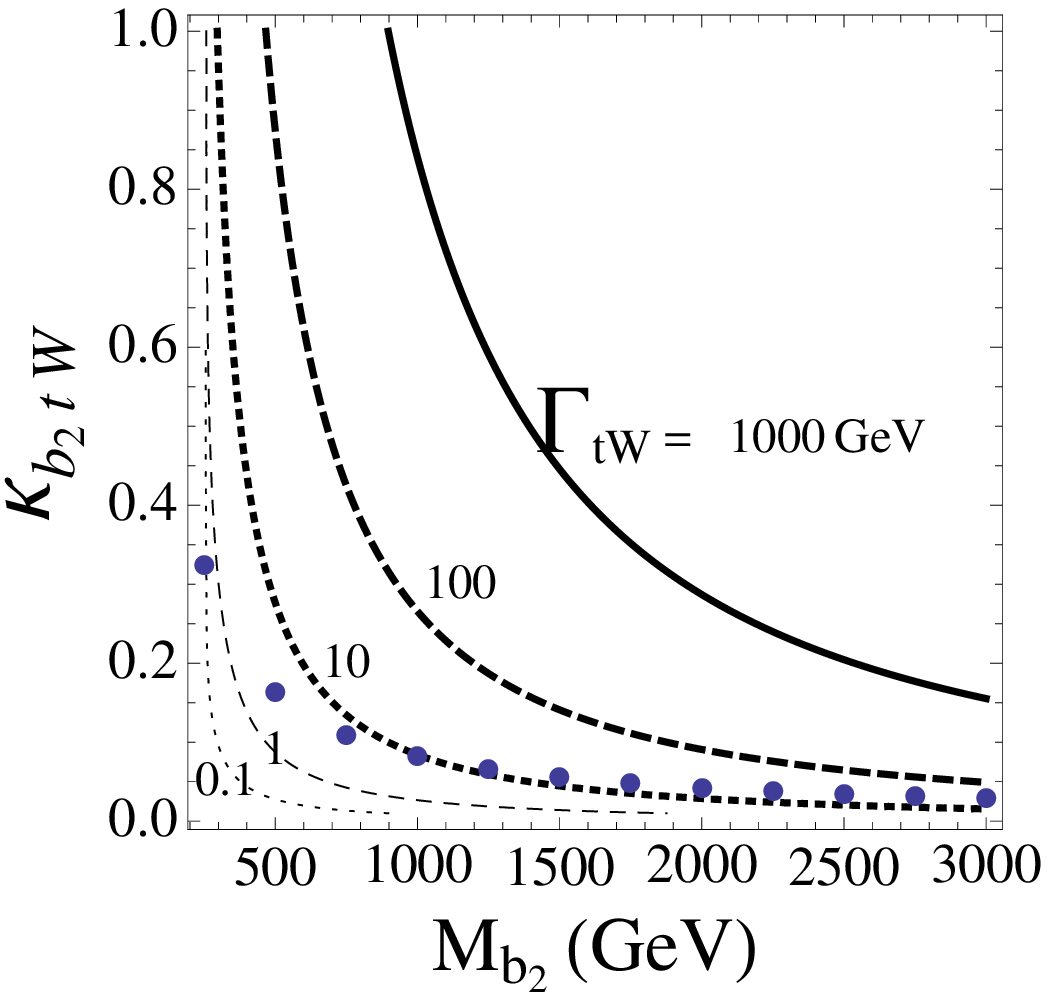}
\includegraphics[width=0.32\textwidth]{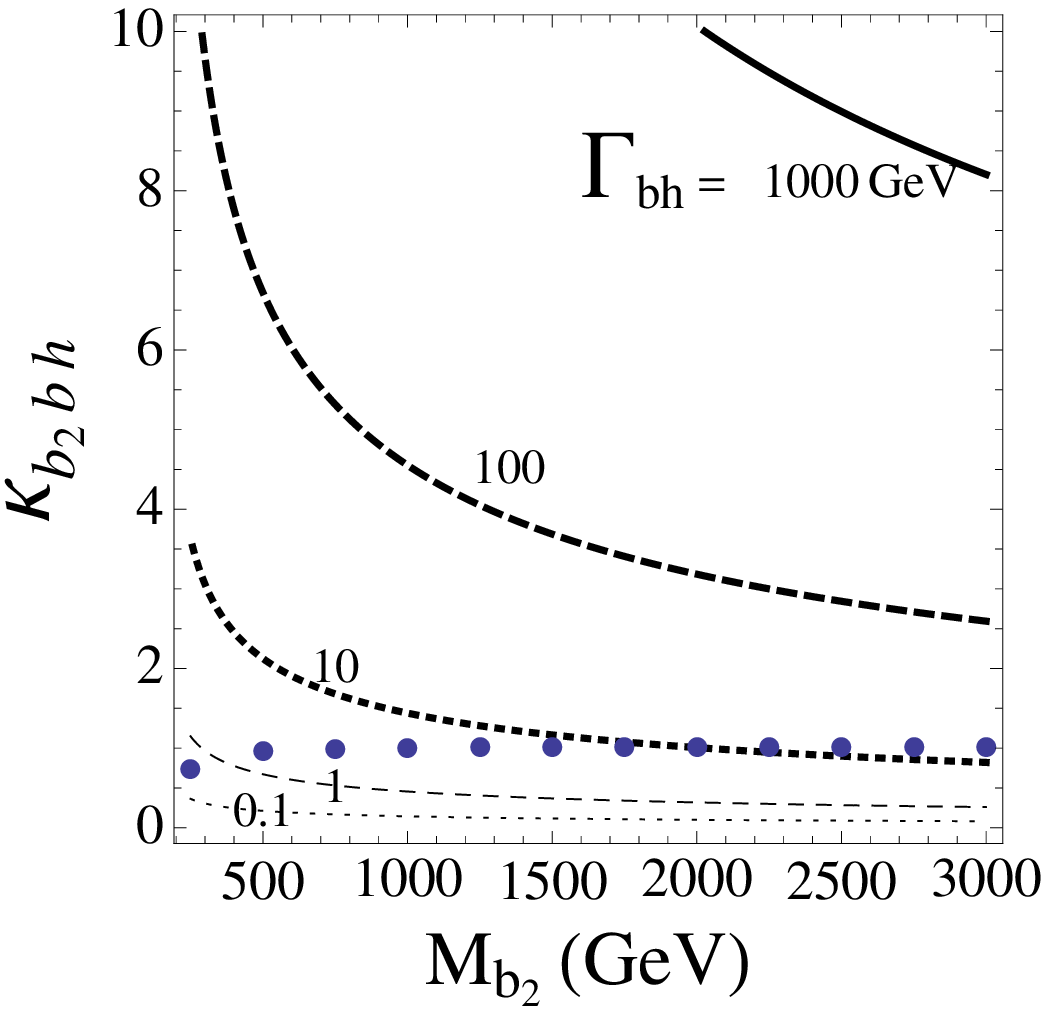}
\caption{Contours of partial-widths of the $b'$ to the $bZ$, $tW$ and $bh$ final states. 
The blue dots show the relation between the $M_{b_2}$ and $\kappa$ as shown in Table~\ref{MbpKappa.TAB}.
\label{b2-GamMI.FIG}}
\end{center}
\end{figure}
%

\section{LHC signatures}
\label{LHCSign.SEC}
At a hadron collider such as the LHC, the production can proceed through the $gg$, $gq$ and $qq$ initial states,
where instead of $q$ we can have a $b$-quark too. 
For sub-TeV $b'$ mass, we expect the $g$ parton distribution function (pdf) to be bigger than the $q$ and $b$ pdf, and therefore 
we expect the $gg$, $gq$ and $qq$ signal (and background) rates to be in decreasing order.
Therefore, to get good significance, if the signal is $qq$ initiated for example, the background should not be 
$gg$ or $gq$ initiated, and similarly for the other possibilities. 
If the $b'$  is not too heavy, the $gg\to b_2 b_2$ pair-production is expected to have the largest production rate compared to 
single production owing to the larger gluon pdf. But the QCD background will also be large. 
For processes for which QCD induced background is not present, the single-production channel can lead to a good reach at the LHC. 
Single production of the $b'$ proceeds via the offdiagonal couplings in Eqs.~(\ref{b'uni.EQ})~and~(\ref{Lagb'H.EQ}). 

In this study, we consider $p p \rightarrow b'\bar{b'}, b' Z~{\rm and}~b' h$ processes
as the discovery channel of the $b'$ and to show its vector-like character. 
We compute the signal cross-section for various masses and compute the main irreducible SM backgrounds
for these channels using Monte Carlo event generators. 
We have defined the warped-space model with the vector-like $b'$ in the matrix-element and event generators
MadGraph 5 Version 1.3.2~\cite{Alwall:2007st} and CalcHEP Version 2.5.6~\cite{CalcHEP}, and all our results in this
section are obtained using these event generators. 
We use CTEQ6~\cite{Pumplin:2002vw} parton distribution functions. 

In order to make the multi-particle phase-space Monte Carlo integration tractable timewise, 
wherever possible, we use the narrow-width approximation and multiply by the appropriate 
branching ratios in order to obtain the required cross-section in the channel considered. 
This will mean that the acceptance in transverse momentum ($p_T$) and rapidity ($y$) for the final state particles will not be taken
into account exactly, but since we mostly deal with high $p_T$ particles, the inaccuracies should be small. 
These agree very well as the $p_T$ of the $Z$ becomes large, 
and we find, for instance, the agreement to be better than 10~\% for $M_{b'} \gtrsim 500~$GeV.

In the following, we analyze $b'$ production at the LHC followed by the 
$b' \to bZ,\, tW,\, {\rm or}\, bh$ decay modes. As mentioned in Sec.~\ref{Intro.SEC}, this
will help in revealing the vector-like nature of the $b'$. 
We devise kinematic cuts to establish signal events above SM background, and obtain the luminosity required
for the benchmark points in Table~\ref{MbpKappa.TAB}
at the 14~TeV LHC to obtain at least $5\, \sigma$ statistical significance and for observing at least 
10 events.

To obtain model-independent results, we use the cross-sections for the benchmark points in Table~\ref{MbpKappa.TAB} and
factor-out the known dependence on the couplings $\kappa$ and $M_{b'}$ to make a fit to the 
purely kinematical part of the cross-section (including the pdf and phase-space factors). Once this fit is made, we fold back in the 
dependence on the couplings and mass and obtain the cross-section for any value of these parameters model-independently,
and infer the required luminosity.\footnote{For a model-independent study in the extended MSSM context, see Ref.~\cite{Kang:2007ib}.}

\subsection{$p p \rightarrow b' \bar{b'}$ process}
In this section we analyze the $b'$ pair production which is initiated by the $gg$ initial state
as shown in Fig.~\ref{gg2bpbp.FIG}.
\begin{figure}[h!]
\begin{center}
\includegraphics[width=0.4\textwidth]{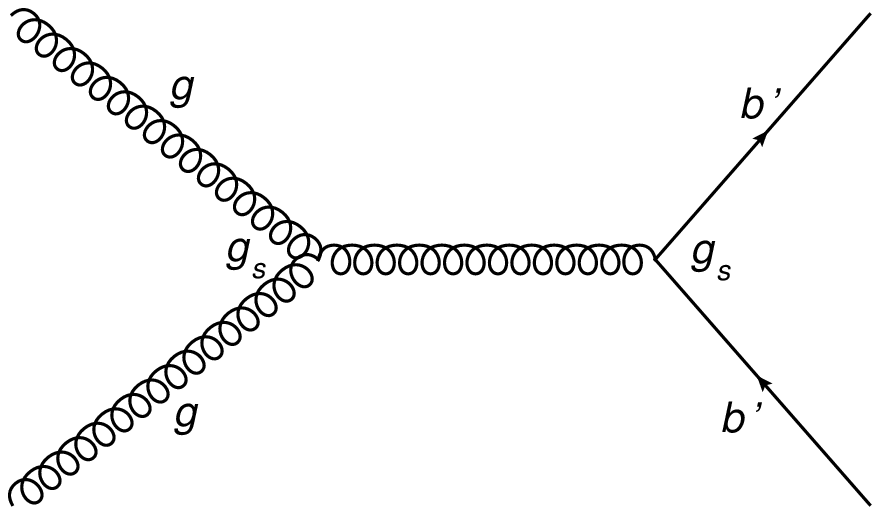}
\hspace*{1cm}
\includegraphics[width=0.25\textwidth]{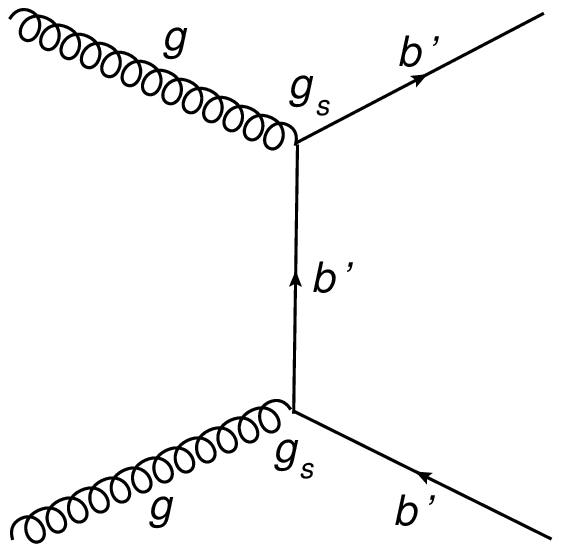}
\caption{The partonic Feynman graphs for $pp\to b'\bar{b'}$ at the LHC.
We show only the $gg$ initiated graphs as examples.
\label{gg2bpbp.FIG}}
\end{center}
\end{figure}
Since the production cross-section is mostly dominated by the $b'$ coupling to the gluon
(with gauge coupling constant $g_S$), our results for the production are largely model-independent.\footnote{We 
have roughly estimated the effective $ggh$ (top triangle diagram) contribution to $b' \bar{b'}$ production and find this 
to be much smaller than the gluon exchange contribution. 
}
In Fig.~\ref{CSBCbpbp-LumibZbZ.FIG} (left) we show the $p p \to b' \bar{b'}$ cross-section in fb as a function of $M_{b_2}$
after $p_T$ and $y$ cuts. 
These cuts are applied after the $bZ$ decay of both the $b'$, requiring $-2.5 < y_{b,Z} < 2.5$ and ${p_T}_{b,Z} > 25~$GeV 
as we detail next. 
\begin{figure}
\begin{center}
\includegraphics[width=0.48\textwidth]{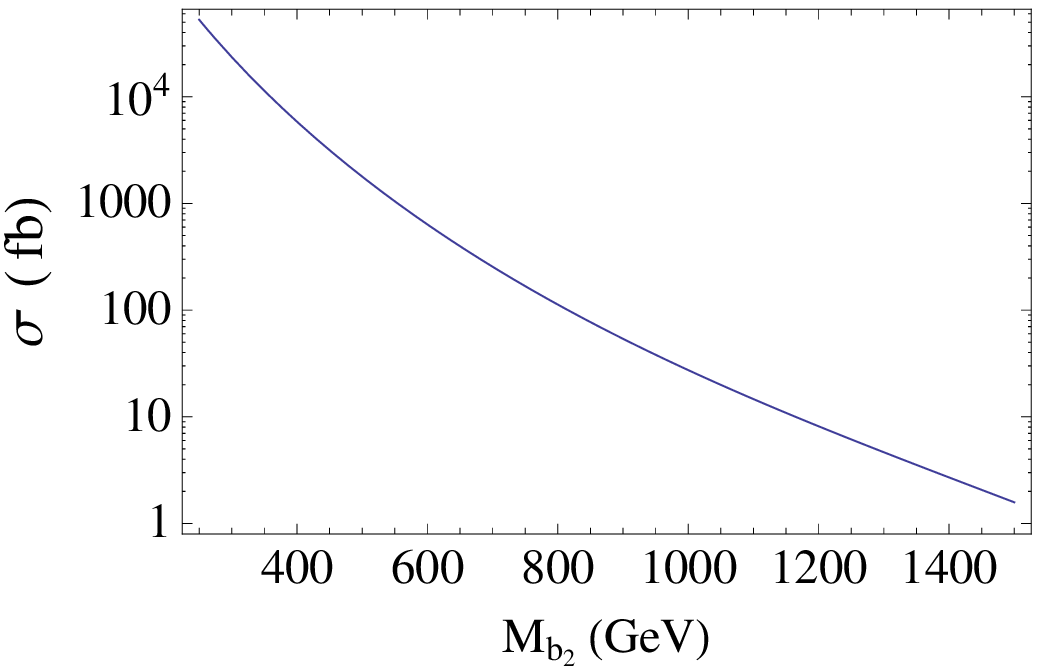}
\hspace*{0.02\textwidth}
\includegraphics[width=0.48\textwidth]{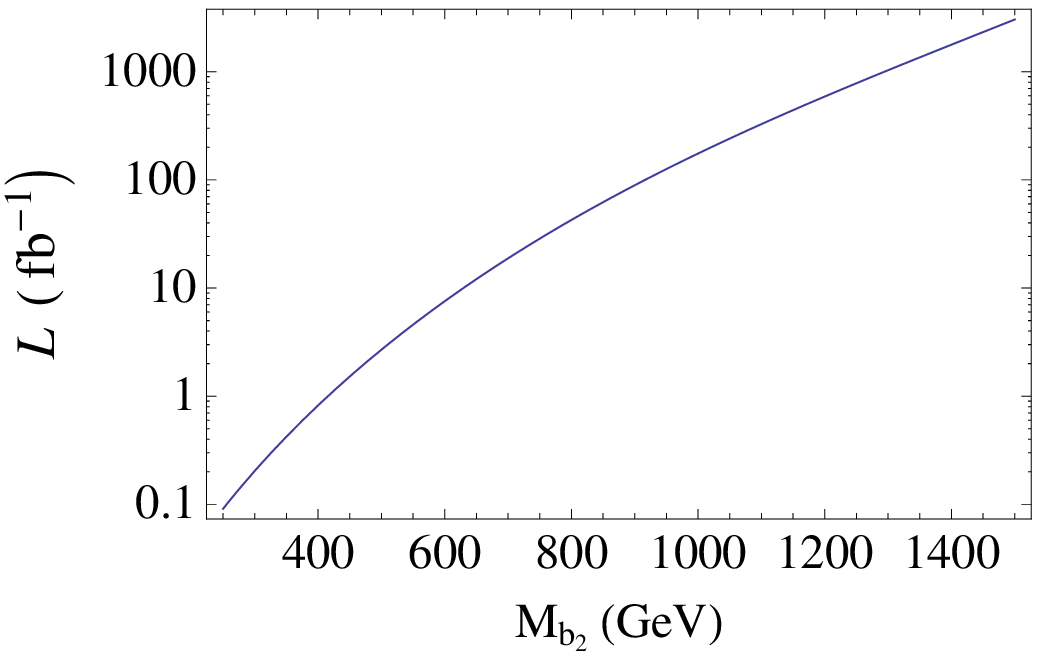}
\caption{$p p \to b' b'$ cross-section after $p_T$ and $y$ cuts (left),
and,
the luminosity-required for $5\, \sigma$ significance with at least 10 signal events 
in the $pp \to b' \bar{b'} \to b Z \bar{b} Z \to b \ell\ell \bar{b} jj$ channel after all cuts (right), 
with $BR(b' \to b Z) = 1/3$ assumed. 
These are for the 14~TeV LHC.  
\label{CSBCbpbp-LumibZbZ.FIG}}
\end{center}
\end{figure}
%

\subsubsection*{$b' \bar{b'} \to bZ \bar{b} Z$ decay mode:}
We consider here both the $b'$s decaying into the $bZ$ final state resulting in the $bZ\bar{b}Z$ final state.
We demand two tagged-$b$s, consider the semileptonic channel taking one of the $Z$s to decay hadronically 
(including only $u,d,c,s$, but not the $b$) and the other
$Z$ decaying leptonically ($\ell = e \ {\rm and}\ \mu$ with $BR(Z\to\ell\ell) = 0.066$), resulting in the channel
$pp \rightarrow b' \bar{b'} \rightarrow bZ \bar{b}Z \rightarrow b l^{+}l^{-} \bar{b}jj$. 
To avoid having to deal with combinatorics issues with the four $b$s that will be present if the
$Z$ decays to $b\bar b$, we ask that this not happen by demanding that the tagged-$b$ 
is not among the two jets that reconstruct to the $Z$. 
We obtain the signal and electroweak background cross-section at the $bZbZ$ level and multiply the $\sigma(pp\to bZbZ)$ cross section by the factor
$2 \eta^2_b\, BR_{Z\to \ell \ell}\, (BR_{Z\to jj} + BR_{Z\to bb}(1-\eta_b)^2) = 0.019$ with $j=\{u,d,c,s\}$, 
where, $\eta_b$ is the $b$-tagging efficiency, the $BR_{Z\to bb}$ term counts the $Z\to b \bar b$ decays that fail the $b$-tag, and a factor 
of 2 is because the hadronic-$Z$ and the leptonic-$Z$ can be exchanged resulting in the same final state. 
We take the $b$-tagging efficiency $\eta_b = 0.5$.
We obtain the QCD background at the $bjjbZ$ level as we explain in more detail below. 

To maximize the signal at the expense of the SM background, we apply the following cuts: \\
\indent \emph{Rapidity}: $-2.5 < y_{b,j,Z} < 2.5$, \\
\indent \emph{Transverse momentum}: ${p_T}_{b,j,Z} > 25~$GeV, \\
\indent \emph{Invariant mass cuts}:  \\
\indent\indent $M_Z - 10~{\rm GeV} < M_{jj} < M_Z + 10~{\rm GeV}$,\\
\indent\indent $0.95 M_{b_2} < M_{(bZ)} < 1.05 M_{b_2}$ . \\ 
where, in the last invariant-mass cut, 
we accept the event if the invariant mass of a $b$ with either $Z$ lies within the invariant mass
window, \emph{and}, the invariant mass of the other b with either $Z$ also lies within the window. 

We show in Table~\ref{pp2bZbZ2bjjbll.TAB} the signal and background cross sections after $y,p_T$ and invariant mass
cuts as a function of $M_{b'}$ 
with the corresponding $\kappa$ as shown in Table~\ref{MbpKappa.TAB},
and show the luminosity required at the 14~TeV LHC for $5\, \sigma$ significance with the requirement
that at least 10 signal events be observed. 
\begin{table}[!h]
\begin{center}
\caption{Signal and background cross-sections at the 14~TeV LHC for the process $p~p\rightarrow b' \bar{b'} \to bZ\bar{b}Z$,
and the luminosity required (${\cal L}$) in the semileptonic decay mode, 
for the benchmark masses and couplings shown in Table~\ref{MbpKappa.TAB}. 
The $bZbZ$ columns do not include $b$-tagging factors, $BR(Z\to\ell\ell)$ or $BR(Z\to jj)$, while $\cal{L}$ includes all these factors.
$(bjjbZ)_{\rm tot}$ shows the total background (including electroweak and QCD) where the QCD background is computed 
using the channels detailed in the second table weighted by appropriate factors as explained in the text.
}
\vspace{0.5cm}
\begin{tabular}{|c|c|c|c|c|c|c|c|} \hline
\multicolumn{1}{|c|}{ } & \multicolumn{2}{c|}{Signal $\sigma_{s}$ (in $fb$)} & \multicolumn{4}{c|}{Background $\sigma_{b}$ (in $fb$)} & \multicolumn{1}{c|}{} \\ \cline{2-7} 
\multicolumn{1}{|c|}{$M_{b_2}$} & \multicolumn{2}{c|}{$bZbZ$} & \multicolumn{2}{c|}{$bZbZ$} & \multicolumn{2}{c|}{$(bjjbZ)_{\rm tot}$} & \multicolumn{1}{c|}{$\mathcal L$} \\ \cline{2-7} 
\multicolumn{1}{|c|}{($GeV$)} & \multicolumn{1}{c|}{$y,p_T$} & \multicolumn{1}{c|}{All} & \multicolumn{1}{c|}{$y,p_T$}& \multicolumn{1}{c|}{All}& \multicolumn{1}{c|}{$y,p_T$}& \multicolumn{1}{c|}{All} & \multicolumn{1}{c|}{($fb^{-1}$) }\\ 
\multicolumn{1}{|c|}{ } & \multicolumn{1}{c|}{cuts} & \multicolumn{1}{c|}{cuts} & \multicolumn{1}{c|}{cuts}& \multicolumn{1}{c|}{cuts}& \multicolumn{1}{c|}{cuts}& \multicolumn{1}{c|}{cuts} & \multicolumn{1}{c|}{ }\\ \hline
250 & 25253 & 25082 & 21.804 &  0.3797 & 16938 & 29.52 & 0.021 \\ \hline
500 & 171.34 & 148.69 & 21.804 &  0.047 & 16938 & 3.74 & 3.514 \\ \hline
750 & 14.508 & 12.221 & 21.804 &  0.0097 & 16938 & 0.997 & 42.752 \\ \hline
1000 & 2.314 & 1.9214 & 21.804 &  0.0027 & 16938 & 0.259 & 271.92 \\ \hline
1250 & 0.484 & 0.399 & 21.804 &  0.0011 & 16938 & 0.048 & 1310 \\ \hline
\end{tabular}
\label{pp2bZbZ2bjjbll.TAB}
\end{center}
\begin{center}
\vspace{0.5cm}
\begin{tabular}{|c|c|c|c|c|c|c|} \hline
\multicolumn{1}{|c|}{} & \multicolumn{6}{c|}{QCD background (in $fb$)} \\ \cline{2-7} 
\multicolumn{1}{|c|}{$M_{b_2}$} & \multicolumn{2}{c|}{$bjjbZ$} & \multicolumn{2}{c|}{$bbjbZ$}& \multicolumn{2}{c|}{$bbbbZ$}\\ \cline{2-7} 
\multicolumn{1}{|c|}{($GeV$)} & \multicolumn{1}{c|}{$y,p_T$}& \multicolumn{1}{c|}{All}& \multicolumn{1}{c|}{$y,p_T$}& \multicolumn{1}{c|}{All} & \multicolumn{1}{c|}{$y,p_T$}& \multicolumn{1}{c|}{All}\\ 
\multicolumn{1}{|c|}{ } & \multicolumn{1}{c|}{cuts}& \multicolumn{1}{c|}{cuts}& \multicolumn{1}{c|}{cuts}& \multicolumn{1}{c|}{cuts}& \multicolumn{1}{c|}{cuts}& \multicolumn{1}{c|}{cuts}\\ \hline
250 & 16790 & 27.304 & 255.41 & 2.7 & 81.01 & 1.92\\ \hline
500 & 16790 & 3.513 &255.41 & 0.256 & 81.01 & 0.194 \\ \hline
750 & 16790 & 0.958 &255.41 & 0.031 & 81.01 & 0.057\\ \hline
1000 & 16790 & 0.2514 &255.41 & 0.0052 & 81.01 & 0.008 \\ \hline
\end{tabular}
\end{center}
\end{table}
The $(bjjbZ)_{\rm tot}$ column in Table~\ref{pp2bZbZ2bjjbll.TAB} shows the total background which is the sum of the QCD and electroweak backgrounds, 
where the QCD background is got from the components shown in the second table as 
$$(bjjbZ)_{\rm QCD} = (bjjbZ) + (1-\eta_b) (bbjbZ) + (1-\eta_b)^2 (bbbbZ) \ , $$ 
where $b$ includes both $b$ and $\bar b$, and the $(1-\eta_b)$ factors take into account a $b$-quark that has failed the $b$-tag, 
i.e. we assume here that a $b$-quark that fails the $b$-tag will be taken to be a light-jet.  
We find that the luminosity required is signal-rate limited.

The results shown here are largely model-independent since the production cross-section mostly relies on the color quantum number of the $b'$ 
since the cross-section is dominated by the gluon exchange contribution, with a coupling of $g_s$. 
In Fig.~\ref{CSBCbpbp-LumibZbZ.FIG} (right) we show the luminosity-required for $5\, \sigma$ significance with 
at least 10 signal events at the 14~TeV LHC, in the $pp \to b' b' \to b Z b Z \to b \ell\ell b jj$ channel after all cuts, 
with $BR(b' \to b Z) = 1/3$ assumed. 

The dileptonic mode, i.e. when both $Z$s decay leptonically, is much cleaner since there is no QCD background, but the BR is smaller. 
Since we are limited by signal rate, we expect the luminosity required to be much bigger than for the semileptonic mode we have
focussed on. The luminosity required for the dileptonic mode can easily be computed from the signal and 
$bZbZ$ background cross-sections given in Table~\ref{pp2bZbZ2bjjbll.TAB} after taking into account the 
$BR_{Z\to \ell\ell}$ for the other $Z$ also. One can also consider demanding only one $b$-tag rather than the two that we have, 
which will increase the signal rate, but so will the background, although the luminosity required may end up being lesser. 

\subsubsection*{$b'\bar{b'} \to bZ \bar{b} h$ and other decay modes:}
We only consider a light Higgs decaying as $h\to b \bar b$ (with $BR \approx 1$), 
i.e. the $b'\bar{b'} \to bZ \bar{b} h \to b Z \bar{b} b \bar b$ channel, and demand four $b$-tags.  
For this, the $\sigma$ multiplied by the branching fractions and $b$-tagging efficiency, shown earlier, will be about half the $bZbZ$ case shown in 
Table~\ref{pp2bZbZ2bjjbll.TAB} and in Fig.~\ref{CSBCbpbp-LumibZbZ.FIG} (left). 
The dominant SM backgrounds will
then be $bbbbZ$, which we have already computed for the previous case and shown in Table~\ref{pp2bZbZ2bjjbll.TAB}. 
As we can see from this, for large $M_{b'}$, the required luminosity will be signal-rate limited as it was in the previous
case, and therefore the luminosity required will be about twice that needed for the $bZbZ$ case shown in Table~\ref{pp2bZbZ2bjjbll.TAB}
and in Fig.~\ref{CSBCbpbp-LumibZbZ.FIG} (right). 

One could also consider the $bZtW$ or other combinations of decay modes of the $b'$ pair, but 
we do not consider these here, as our main motivation is to focus on those decay-modes which help in revealing
aspects of the vector-like nature of the $b'$.

\subsection{$p p \rightarrow b' Z, b' h$ processes}

In this section, we analyze the $pp \to b' Z$ and $pp \to b' h$ processes which are initiated by the $bg$ initial state
as shown in Fig.~\ref{bg2bpZ-h.FIG}.
\begin{figure}[h!]
\begin{center}
\includegraphics[width=0.4\textwidth]{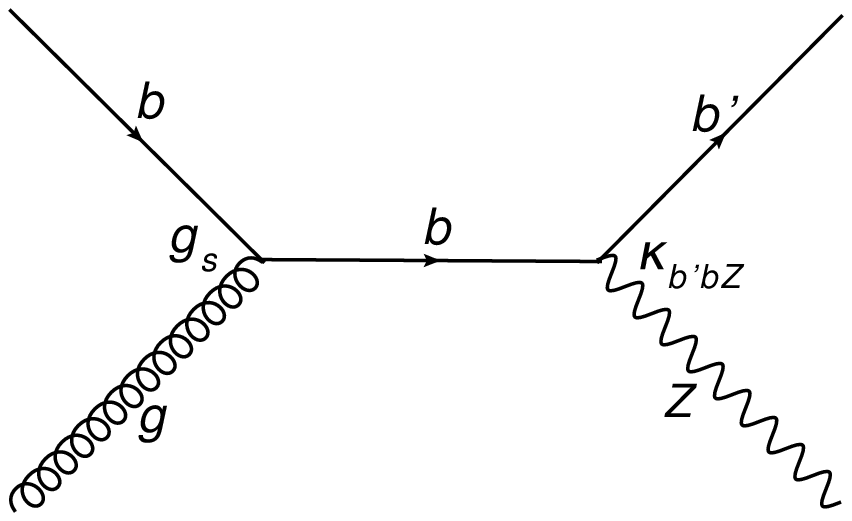}
\hspace*{1cm}
\includegraphics[width=0.4\textwidth]{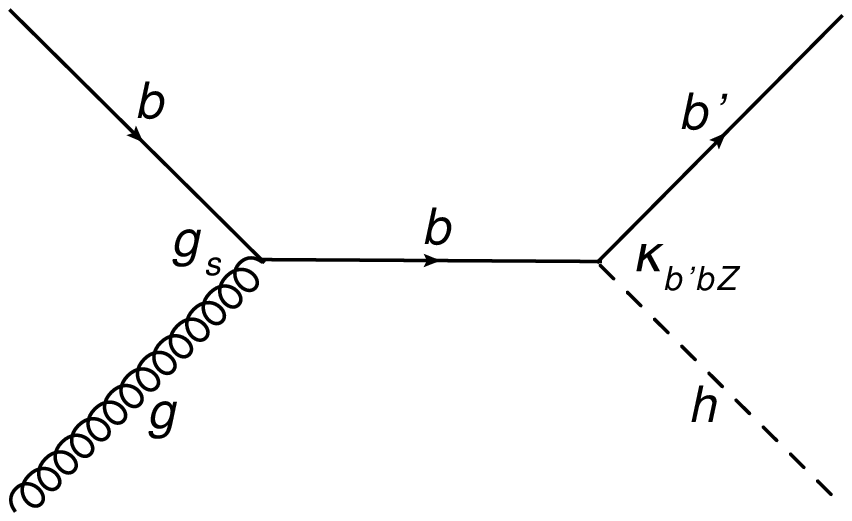}
\caption{The partonic Feynman graphs for $pp\to b'Z, b'h$ at the LHC.
\label{bg2bpZ-h.FIG}}
\end{center}
\end{figure}
In Fig.~\ref{CSBCbpZ-LumibZZ.FIG} (left) we show contours of the $p p \to b' Z$ cross-section in $fb$, after $p_T$ and $y$ cuts, 
in the $\kappa^{L}_{{b_2}bZ}$~--~$M_{b_2}$ plane at the 14~TeV LHC. 
These cuts are applied after the $b'\to bZ$ decay, requiring $-2.5 < y_{b,Z} < 2.5$ and ${p_T}_{b,Z} > 0.1 M_{b_2}$. 
The blue dots show the $M_{b'}$ and $\kappa^L_{b_2 b Z}$ as given in Table~\ref{MbpKappa.TAB}.
\begin{figure}[h]
\begin{center}
\includegraphics[width=0.49\textwidth]{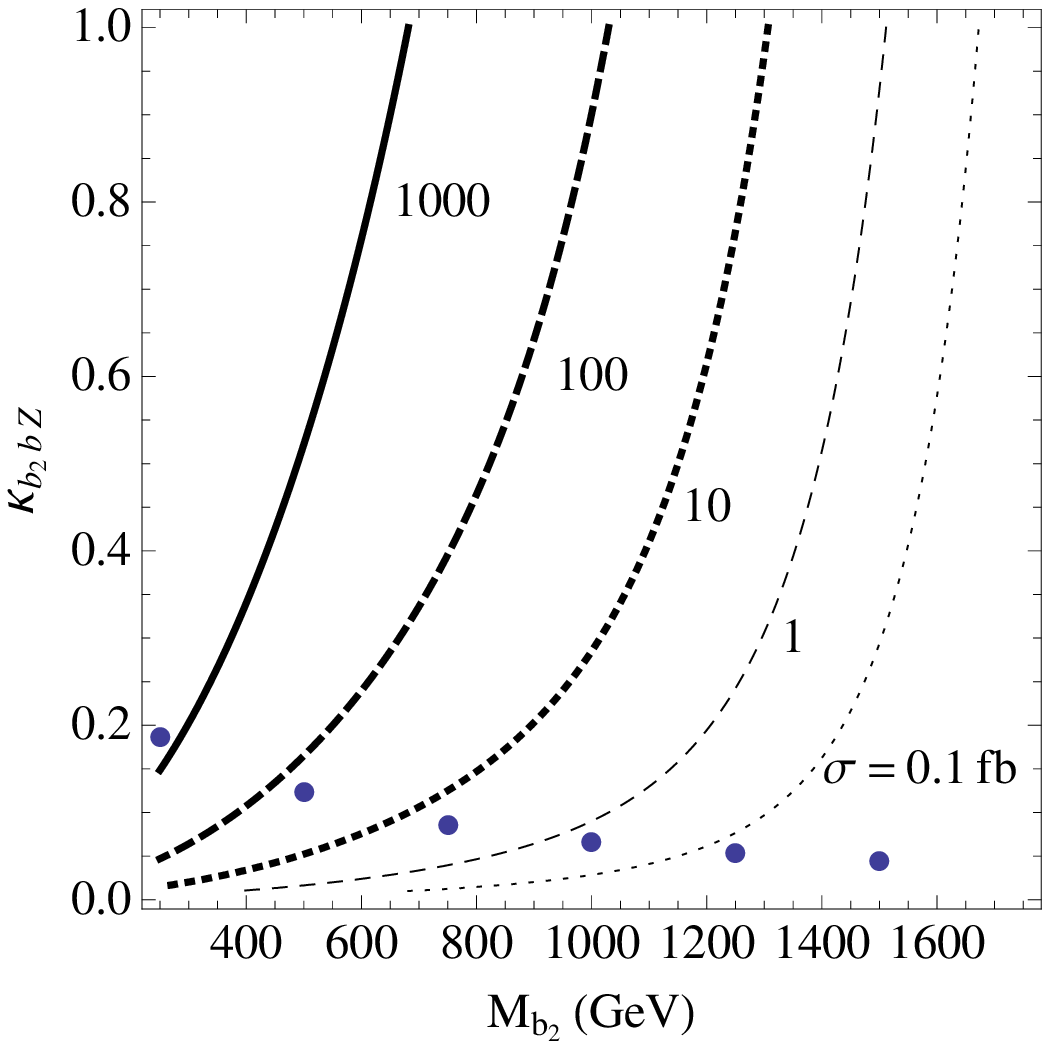}
\includegraphics[width=0.49\textwidth]{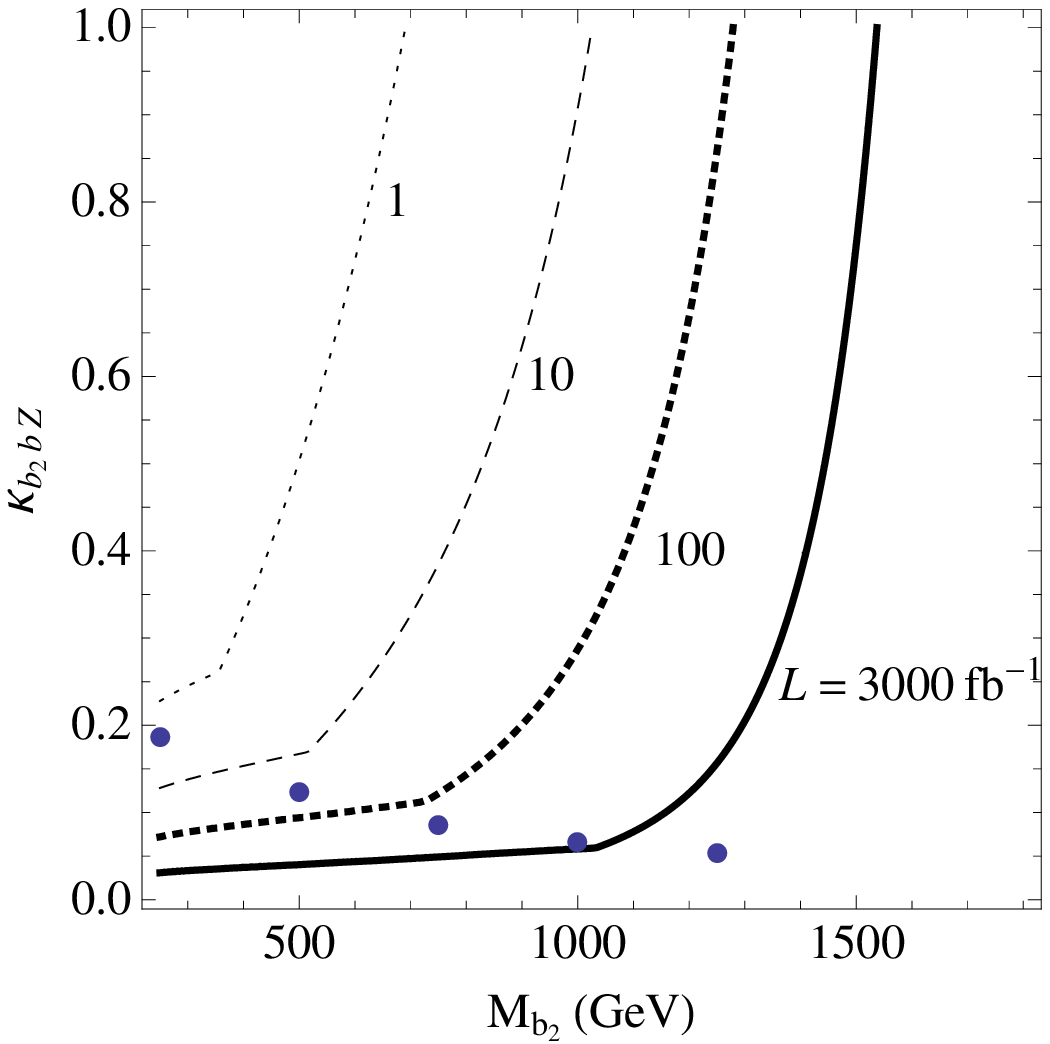}
\caption{Model-independent contours of the $p p \to b' Z$ cross-section in fb after $p_T$ and $y$ cuts (left),
and, contours of the luminosity-required for $5\, \sigma$ significance with at least 10 signal events in the 
$pp \to b' Z \to b Z Z \to b \ell\ell jj$ channel after all cuts (right), 
with the region to the left of a contour covered by that luminosity, and $BR(b' \to b Z) = 1/3$ assumed.  
These are for the 14~TeV LHC.  
The blue dots show the $M_{b'}$ and $\kappa_{b_2 b Z}$ as given in Table~\ref{MbpKappa.TAB}.
\label{CSBCbpZ-LumibZZ.FIG}}
\end{center}
\end{figure}

The $b' h$ cross-section is expected to be similar to the $b g \rightarrow b' Z$ case above. 
In the following, we consider the $b' \to bZ,\, tW,\, {\rm or}\, bh$ decay modes. 
For the $b Z h$ final state
both $b g \to b' h \to b Z h$, and $b g \to b' Z \to b h Z$ channels 
will contribute. 
We will discuss each of these channels, in turn, next.  

\subsubsection*{$bg \to b' Z \to bZZ$ channel:}
We will consider next, in turn, the semileptonic decay mode when one $Z$ decays leptonically and the other hadronically (i.e. $bZZ \to bjj\ell^+\ell^-$), and, 
dileptonic decay mode when both $Z$s decay leptonically (i.e. $bZZ \to b\ell^+\ell^-\ell^+\ell^-$). 

\medskip
\noindent \underline{\emph{Semileptonic decay mode}}:
For the semileptonic $p p \to b' Z \to b Z Z \to b jj \ell^+ \ell^-$ channel,
we assume that the leptonically decaying $Z$ is fully reconstructed, and perform our analysis at the $bjjZ$ level. 
We multiply the cross-section at the $bjjZ$ level by $BR(Z\to \ell\ell)$. 
We could have indeed performed the analysis at the $bZZ$ level, but because this channel will be limited by
QCD background as we demonstrate below, we include the latter and perform the analysis at the $bjjZ$ level. 
We demand one tagged $b$-jet, and apply the following cuts:\\
\indent\emph{Rapidity}: $-2.5 < y_{b,j,Z} < 2.5$,\\
\indent\emph{Transverse momentum}: ${p_T}_{b,j,Z} > 0.1 M_{b_2}$ ,\\
\indent\emph{Invariant mass cuts}: \\ 
\indent\indent $M_Z - 10~{\rm GeV} < M_{jj} < M_Z + 10~{\rm GeV}$,\\
\indent\indent $0.95 M_{b_2} < M_{(bZ)\ OR\ (bjj)} < 1.05 M_{b_2}$,\\
where $Z$ means the leptonically decaying $Z$, and in the last invariant mass cut we accept the event if either of $M_{bZ}$ \emph{OR} $M_{bjj}$ lies within 
the window.  
Here, $j$ will exclude the $b$ to avoid having to deal with combinatorics issues with the three $b$s that will be present if the
$Z$ decays to $b\bar b$. We ask that this not happen by demanding that the tagged-$b$ 
is not among the two jets that reconstruct to the $Z$. 
We therefore multiply the signal $bjjZ$ and electroweak background $(bjjZ)_{EW}$ cross sections by 
$\eta_b\, BR_{Z\to \ell \ell} = 0.033$
with $j=\{u,d,c,s\}$, 
where, we include 
the $Z\to b \bar b$ decays that fail the $b$-tag.
Since experimentally light-quark jets and gluon jets cannot be differentiated effectively, 
for the background, we take $j=\{g,u,d,c,s\}$, 
and in addition to the $bZZ$ SM background for which the multiplicative factor is as shown above,
 we include the QCD backgrounds, namely, 
$$ (bjjZ)_{QCD} = (bjjZ) + (bjbZ) (1-\eta_b) + (bbbZ) (1-\eta_b)^2 ,$$
where a $(1-\eta_b)$ factor is included for a $b$-quark that fails to be tagged, and, 
we multiply these with an overall multiplicative factor of $\eta_b\, BR_{Z\to \ell \ell}$.  

The signal and the background cross-sections along with the 
luminosity required for the semileptonic decay mode
for various values of $M_{b'}$ and $\kappa$ given in Table~\ref{MbpKappa.TAB}
are shown in Table~\ref{pp2bzzSL.TAB}.  
In the table, \emph{primary cuts} includes all cuts except for the $M_{(bZ)\ OR\ (bjj)}$ invariant mass cut.  
\begin{table}[h]
\caption{Signal and background cross-sections at the 14TeV LHC for the $pp\rightarrow b' Z \to bZZ \to bjjZ$ channel with its 
charge-conjugate process also included. 
The luminosity required in shown for the semileptonic decay modes 
corresponding to the benchmark masses and couplings shown in Table~\ref{MbpKappa.TAB}. 
The $bjjZ$ columns neither include $b$-tagging factors nor $BR(Z\to\ell\ell)$, while $\cal{L}_{\rm{SemiLep}}$ is shown after all these factors are included.
$(bjjZ)_{QCD}$ shows the total QCD background computed using the 
different channels detailed in the second table weighted by appropriate factors as explained in the text.
}
\begin{center}
\begin{tabular}{|r|r|r|r|r|r|r|r|} \hline
\multicolumn{1}{|c|}{} & \multicolumn{2}{c|}{signal $\sigma_{s}$ (in fb)} & \multicolumn{4}{c|}{background $\sigma_{b}$ (in fb)} & \multicolumn{1}{c|}{} \\ \cline{2-7}
\multicolumn{1}{|c|}{$M_{b'}$} & \multicolumn{2}{c|}{$bjjZ$} & \multicolumn{2}{c|}{$(bjjZ)_{EW}$} & \multicolumn{2}{c|}{$(bjjZ)_{\rm{QCD}}$} & \multicolumn{1}{c|}{$\cal{L}_{\rm{SemiLep}}$} \\ \cline{2-7}
\multicolumn{1}{|c|}{(GeV)} & \multicolumn{1}{c|}{Primary} & \multicolumn{1}{c|}{all} & \multicolumn{1}{c|}{Primary} & \multicolumn{1}{c|}{all} & \multicolumn{1}{c|}{Primary} & \multicolumn{1}{c|}{all} & \multicolumn{1}{c|}{$(fb^{-1})$} \\ 
\multicolumn{1}{|c|}{} & \multicolumn{1}{c|}{cuts} & \multicolumn{1}{c|}{cuts} & \multicolumn{1}{c|}{cuts} & \multicolumn{1}{c|}{cuts} & \multicolumn{1}{c|}{cuts} & \multicolumn{1}{c|}{cuts} & \multicolumn{1}{c|}{} \\ \hline
250 & 1017.66 & 995.86 & 77.03 & 10.33 & 7853.02 & 867.82 & 0.66 \\ \hline
500 & 16.84 & 15.50 & 8.81 & 0.68 & 419.75 & 14.11 & 45.94 \\ \hline
750 & 1.26 & 1.14 & 1.85 & 0.10 & 56.26 & 0.86 & 551.26 \\ \hline
1000 & 0.14 & 0.12 & 0.47 & 0.01 & 12.38 & 0.05 & 3399.67 \\ \hline
\end{tabular}
\begin{tabular}{|r|r|r|r|} \hline
\multicolumn{1}{|c|}{$M_{b'}$} & \multicolumn{3}{c|}{QCD background (in fb)} \\ \cline{2-4} 
\multicolumn{1}{|c|}{(GeV)} & \multicolumn{1}{c|}{$bjjZ$} & \multicolumn{1}{c|}{$bjbZ$} & \multicolumn{1}{c|}{$bbbZ$} \\ \hline
250 & 546.36 & 634.32 & 17.19 \\ \hline
500 & 10.14 & 7.76 & 0.35 \\ \hline
750 & 0.52 & 0.66 & 0.03 \\ \hline
1000 & 0.02 & 0.06 & 0.002 \\ \hline
\end{tabular}
\end{center}
\label{pp2bzzSL.TAB}
\end{table}
The required luminosity for the semileptonic case is always background limited.

In Fig.~\ref{CSBCbpZ-LumibZZ.FIG} (right) we show the model-independent contours of the 14~TeV LHC luminosity-required for 
$5\, \sigma$ significance with at least 10 signal events in the $\kappa_{b_2bZ}$ -- $M_{b_2}$ plane. 
The region to the left of a contour is covered by that luminosity.
$BR(b' \to b Z) = 1/3$ is assumed.
The kink seen is the cross-over from being background-limited at lower masses to signal-rate-limited at higher masses.   
The blue dots show the $M_{b'}$ and $\kappa_{b_2 b Z}$ given in Table~\ref{MbpKappa.TAB} for which Table~\ref{pp2bzzSL.TAB} applies.

\medskip
\noindent \underline{\emph{Dileptonic decay mode}}:
For the $p p \to b' Z \to b Z Z \to b \ell^+ \ell^- \ell^+ \ell^-$, we perform the analysis at the $bZZ$ level and multiply 
the cross-section by $\eta_b * BR(Z\to\ell\ell)^2$. 
We apply the following cuts:

\indent\emph{Rapidity}: $-2.5 < y_{b,Z} < 2.5$,\\
\indent\emph{Transverse momentum}: ${p_T}_{b,Z} > 25~$GeV,\\
\indent\emph{Invariant mass cut}:  
  $0.95 M_{b_2} < M_{(bZ)} < 1.05 M_{b_2}$,\\
where $Z$ means either of the leptonically decaying $Z$,
and in the invariant mass cut, $M_{bZ}$ is evaluated for both the $Z$s with the event kept if either one of them falls within the window.
We have relaxed the $p_T$ cut here since we do not have to suppress the largish QCD background that we had to contend with in the semileptonic case. 
The signal and background cross-sections along with the 
luminosity required for the dileptonic decay mode
for various values of $M_{b'}$ and $\kappa$ given in Table~\ref{MbpKappa.TAB}
are shown in Table~\ref{pp2bzzDL.TAB}.  
As before, in the table, \emph{primary cuts} includes all cuts except for the $M_{(bZ)}$ invariant mass cut.  
\begin{table}[h]
\caption{Signal and background cross-sections at the 14TeV LHC for the $pp\rightarrow b' Z \to bZZ$ with its 
charge-conjugate process also included, and the luminosity required for the dileptonic decay mode corresponding to the benchmark masses and couplings 
shown in Table~\ref{MbpKappa.TAB}. 
The $bZZ$ columns neither include $b$-tagging factors nor $BR(Z\to\ell\ell)$, while $\cal{L}_{\rm{DiLep}}$ includes all these factors.
}
\begin{center}
\begin{tabular}{|r|r|r|r|r|r|} \hline
\multicolumn{1}{|c|}{} & \multicolumn{2}{c|}{signal $\sigma_{s}$ (in fb)} & \multicolumn{2}{c|}{background $\sigma_{b}$ (in fb)} & \multicolumn{1}{c|}{}\\ \cline{2-5}
\multicolumn{1}{|c|}{$M_{b'}$} & \multicolumn{2}{c|}{$bZZ$} & \multicolumn{2}{c|}{$bZZ$} & \multicolumn{1}{c|}{$\cal{L}_{\rm{DiLep}}$} \\ \cline{2-5}
\multicolumn{1}{|c|}{(GeV)} & \multicolumn{1}{c|}{Primary} & \multicolumn{1}{c|}{All} & \multicolumn{1}{c|}{Primary} & \multicolumn{1}{c|}{All} & \multicolumn{1}{c|}{$(fb^{-1})$} \\ 
\multicolumn{1}{|c|}{} & \multicolumn{1}{c|}{cuts} & \multicolumn{1}{c|}{cuts} & \multicolumn{1}{c|}{cuts} & \multicolumn{1}{c|}{cuts} & \multicolumn{1}{c|}{} \\ \hline
250 & 1119.42 & 1088.84 & 77 & 10.54 & 2.1 \\ \hline
500 & 25.15 & 22.80 & 77 & 2.16 & 97.6 \\ \hline
750 & 2.32 & 2.04 & 77 & 0.52 & 1091.9 \\ \hline
1000 & 0.36 & 0.32 & 77 & 0.15 & 6962.4 \\ \hline
\end{tabular}
\end{center}
\label{pp2bzzDL.TAB}
\end{table}
The required luminosity for the dileptonic case is always signal limited.

\subsubsection*{$bg \to b' Z \to tWZ$ channel:}

In this case, at the $t W Z$ level, the three particles in the final state are different, and therefore there is no combinatorial issue.
For the semileptonic decay mode we have two possibilities, namely, 
when the $Z$ decays leptonically and the W hadronically, and vice-versa. 
If the $Z$ decays hadronically and the $W$ leptonically, we have a neutrino in the final state,
leading to missing momentum. At a hadron machine, since the
incoming parton energies are not known, this missing momentum will prevent the full reconstruction of the
event, but can only be done in the transverse plane. However, one can apply the 
$W$ mass constraint in order to infer ${p_\nu}_z$ (upto a two-fold ambiguity) as explained, for example, in 
Ref.~\cite{Gopalakrishna:2010xm}.

The signal and SM background at the $t W Z$ level are shown in Table~\ref{pp2btw.TAB}.
The choice for all the cuts here is similar to the ones for the dileptonic $bZZ$ case above.
\begin{table}[h]
\caption{Signal and background cross-sections for the $pp \to b' Z \rightarrow tWZ$ channel with the charge-conjugate process also included. 
The $\kappa$ are taken to be as given in Table~\ref{MbpKappa.TAB}.   
}
\begin{center}
\begin{tabular}{|r|r|r|r|r|} \hline
\multicolumn{1}{|c|}{$M_{b'}$} & \multicolumn{2}{c|}{signal $\sigma_{s}$ (in fb)} & \multicolumn{2}{c|}{background $\sigma_{b}$ (in fb)}  \\ \cline{2-5} 
\multicolumn{1}{|c|}{(GeV)} & \multicolumn{1}{c|}{$y,p_T$ cuts} & \multicolumn{1}{c|}{All cuts} & \multicolumn{1}{c|}{$y,p_T$ cuts} & \multicolumn{1}{c|}{All cuts} \\ \hline
300 & 307.92 & 288.04 & 72.78 & 9.10 \\ \hline
500 & 40.02 & 35.88 & 72.78 & 5.72 \\ \hline
750 & 4.20 & 3.74 & 72.78 & 1.84  \\ \hline
1000 & 0.70 & 0.62 & 72.78 & 0.64 \\ \hline
\end{tabular}
\end{center}
\label{pp2btw.TAB}
\end{table}
Since the $t W$ decay mode is present for a chiral $b'$ also, and 
our main motivation in this study is to expose the vector-like nature of the $b'$, we have not computed the QCD background for this process,
and have not determined the luminosity required. 

\subsubsection*{$bg \to\  b' Z,b' h \ \to b Z h$ channel:}
We assume a light Higgs 
with $h\to b \bar b$ (with $BR \approx 1$), and the $Z$ decaying leptonically,
resulting in the $b\ell^+\ell^-b\bar{b}$ channel. 
We demand three $b$-tags. 
We perform the analysis at the $bZh$ level and multiply the cross-section by $\eta_b^3 * BR(Z\to \ell\ell)$, but for the QCD background which we take 
at the $bZb\bar{b}$ level (multiplied by effectively the same factor). 
The $bZb\bar{b}$ background is the same as in the previous case given in Table~\ref{pp2bzzSL.TAB}.  
We show in Table~\ref{pptobzh.TAB} the signal and background cross-sections and the luminosity required.  
\begin{table}[h]
\caption{Signal and background cross-sections for the leptonic $pp\to b'Z + b'h\to bZh$ channel.
The $bZh$ and $bbbZ$ columns neither include $b$-tagging factors nor $BR(Z\to\ell\ell)$, while $\cal{L}$ includes all these factors.
The $\kappa$ are taken to be as given in Table~\ref{MbpKappa.TAB}.}
\begin{center}
\begin{tabular}{|r|r|r|r|r|r|r|r|} \hline
\multicolumn{1}{|c|}{} & \multicolumn{2}{c|}{signal $\sigma_{s}$ (in fb)} & \multicolumn{4}{c|}{background $\sigma_{b}$ (in fb)} & \multicolumn{1}{c|}{} \\ \cline{2-7}
\multicolumn{1}{|c|}{$M_{b'}$} & \multicolumn{2}{c|}{$bZh$} & \multicolumn{2}{c|}{$bZh$} & \multicolumn{2}{c|}{$bbbZ$} & \multicolumn{1}{c|}{$\cal{L}$} \\ \cline{2-7}
\multicolumn{1}{|c|}{(GeV)} & \multicolumn{1}{c|}{$y,p_T$} & \multicolumn{1}{c|}{All} & \multicolumn{1}{c|}{$y,p_T$} & \multicolumn{1}{c|}{All} & \multicolumn{1}{c|}{$y,p_T$} & \multicolumn{1}{c|}{All} & \multicolumn{1}{c|}{$(fb^{-1})$} \\ 
\multicolumn{1}{|c|}{} & \multicolumn{1}{c|}{cuts} & \multicolumn{1}{c|}{cuts} & \multicolumn{1}{c|}{cuts} & \multicolumn{1}{c|}{cuts} & \multicolumn{1}{c|}{cuts} & \multicolumn{1}{c|}{cuts} & \multicolumn{1}{c|}{} \\ \hline
250 & 1093.10 & 1056.96 & 4.68 & 0.74 & 569.35 & 18.01 & 1.13 \\ \hline
500 & 44.30 & 34.70 & 4.68 & 0.14 & 569.35 & 2.22 & 34.41 \\ \hline
750 & 5.94 & 3.54 & 4.68 & 0.03 & 569.35 & 0.37 & 337.30 \\ \hline
1000 & 1.44 & 0.58 & 4.68 & 0.01 & 569.35 & 0.03 & 2058.67 \\ \hline
\end{tabular}
\end{center}
\label{pptobzh.TAB}
\end{table}
The luminosity is signal-rate limited.

We could perhaps gain in luminosity by only demanding one or two $b$-tags as opposed to the
three we demand here, but then the QCD background may be too large.  
One could also consider the hadronic decay of the $Z$ resulting in the $bbbjj$ channel, but the QCD background may be large. 
We have not considered these here. 

\subsection{Other Processes}
Here we collect some processes that we have considered, but have not analyzed in full detail, since 
based on rough estimates we think that they may lead to a larger luminosity requirement
compared to the ones we have considered in detail above. 
We give below some indication for what cross-sections we expect for these processes for the benchmark points
given in Table~\ref{MbpKappa.TAB}. 

\medskip
\noindent{\bf $b q \to b' q$ process:}
For the process $b q \rightarrow b' q$, the signal is induced by the t-channel exchange of a $Z$. 
We find the signal cross-section to be small compared to the SM background.
For example, for $M_{b_2} = 750$~GeV, the signal cross-section for $b Q \to b' q \to b Z q \to b \ell \ell q$ is about $0.65$~fb, 
which is about 40 times smaller than the background,
which we have computed 
with an invariant mass cut of $|M_{bZ} - M_{b_2}| \leq 25$~GeV. 

\medskip
\noindent{\bf $b q  \rightarrow q b' W, q b' Z, q b' h$ and $b g \rightarrow g b' Z, g b' h$ processes:} 
The channels with a $q$ in the final state proceed via $b W$ and $b Z$ fusion. The backgrounds are also $b W$ and $b Z$ initiated, 
and is potentially under control. 
But since the initial state is only $q$ and $b$, this may not compare well to $g$ initiated processes. 
The background is particularly small for $b q  \to q b' Z \to q b h Z$ since $h$ has to attach to $b$ line
which is suppressed by $\lambda_b$, the $b$-quark Yukawa coupling, and there's no $ZZh$ coupling. 
Similar situation should also apply for the channel $b q  \to q b' h \to q b h h$. 
Since experimentally we cannot tell the difference between a light $q$ and $g$, we should include 
$b g \rightarrow g b' Z, g b' h$ here, which will result in the same final state as the above processes. 

We expect these 3-body final state processes in general to have smaller cross-section compared to the 2-body final states
considered earlier. 
For $M_{b_2} = 750\, GeV$ and $b'$ decaying as $b' \rightarrow b Z$ the total signal strength is about $0.08$~fb
(which includes the charge conjugate process), with 
one of the Z decaying leptonically and the other decaying into light jets.

\medskip
\noindent{\bf $q g \rightarrow q b' b, q b' t$ processes:}
These proceed via $g Z$ and $g W$ fusion respectively. 
Comparing to the $b g \to b' Z$ process, we see that this is a 3-body final state which would suppress the cross-section. 

For $b' \to b h$, the $q b h b$ irreducible background should be small since it is suppressed by $\lambda_b^2$.
But, the SM background will include processes in which the $q$ is replaced by a $g$, which will mean that the background
is $gg$ initiated, and is likely to be much larger.

\medskip
\noindent{\bf $g g \rightarrow b' b Z, b' t W, b' b h$ processes:}
These processes are related to the $g g \to b' b'$, and being a 2-body process, it will be clearly bigger than the above 3-body processes
if the $M_{bZ,tW,bh} = M_{b'}$ region is included.
These channels will be important only if $M_{b_2}$ is so large that phase-space considerations will 
favor this channel over on shell pair-production. 

\medskip
\noindent{\bf $q q \rightarrow b' b, b' t$ processes:}
The signal for the $b' b$ final state is small as this is a $qq$ initiated process. 
For example, if we consider the $b'$ decaying into a $b$ and a $Z$
with the $Z$ decaying leptonically, the signal turns about $0.009$~fb for $M_{b_2} = 750$~GeV.
Moreover, the background, which has $gg$ initiated contributions, is expected to be much bigger than the signal. 

\medskip
\noindent{\bf $g g \to b' b$ and $g b \to b' g$ process:}
These proceed via s-channel and t-channel Higgs exchange respectively, with an effective 
$ggh$ vertex (top triangle diagram). 
We roughly estimate this contribution to be potentially bigger than the $\sigma(b g \to b' Z)$
we have considered earlier; however these channels are susceptible to the $gg$ initiated SM background 
which is large, and therefore might lead to a larger required-luminosity.

\section{Conclusions}
\label{CONCL.SEC}
Many beyond the standard model extensions predict the existence of heavy vector-like fermions.  
We consider the phenomenology of one such vector-like fermion, called $b'$, with electromagnetic charge $-1/3$
in a model-independent fashion. 
We write a general Lagrangian containing interactions of the $b'$ with SM fields,
identify the relevant parameters, namely the $b'$ mass, and, $b'bZ$, $b'tW$ and $b'bh$ couplings. 
We present analytical expressions for the $b'$ partial widths to $tW$, $bZ$ and $bh$ final states. 

Our main focus is the LHC signatures of a vector-like $b'$, the characteristic of which is the ${\cal O}(1)$
branching ratio into the $bZ$ and $bh$ decay modes in addition to the $tW$ mode which is also present for
a chiral (fourth generation) $b'$. 
Since our goal is to expose aspects unique to a vector-like $b'$ 
we consider the former two decay modes in detail. 

We explore the $b'\bar{b'}$ pair production and, $b'Z$ and $b' h$ single production processes at the 14~TeV LHC followed by their decays
as mentioned above, namely, $b'\bar{b'} \to bZ \bar{b}Z$, $b'\bar{b'} \to bZ \bar{b}h$, $b'Z \to bZZ$ and $b'Z + b' h \to bZh$ channels. 
We list a few other $b'$ single production processes very briefly and mention the reasons why we do not consider them in detail.  
For the modes with two $Z$s, we consider the semileptonic decay mode where one $Z$ decays hadronically and the other leptonically,
and the dileptonic mode where both $Z$s decay leptonically, while, for the modes with a Higgs, we only consider the semileptonic
mode where the $Z$ decays leptonically and the Higgs into $b\bar b$ which is valid for a light Higgs. 

We compute signal and background cross-sections after $pT$, rapidity and invariant mass cuts.
As $M_{b'}$ goes from 250~GeV to 1~TeV, for the benchmark couplings shown in Table~\ref{MbpKappa.TAB}, 
the $b'\bar{b'}$ pair production signal cross-section after our cuts ranges from about 68~pb to 28~fb, 
while the $b'Z + b' h$ single production cross-section ranges from about 1.4~pb to 0.4~fb.
These are after including the corresponding charge-conjugate processes. 
We also show model-independent plots for how these cross-sections vary as the $b'tW$, $b'bZ$, and $b'bh$ couplings and $M_{b'}$ vary. 
We identify the dominant SM backgrounds for the semileptonic and dileptonic decay modes, including the dijet QCD background for the
semileptonic mode, in addition to the irreducible electroweak background. The dijet QCD background is substantial. 
For the $b'\bar{b'}\to bZbZ\to bjjb\ell^+\ell^-$ channel we find the LHC reach to be about $M_{b'} \approx 1250~$GeV with about $1300~fb^{-1}$,
while for the $b'Z+b'h \to bZh \to b\ell^+\ell^-b\bar{b}$ channel it is about $M_{b'} \approx 1000~$GeV with about $2050~fb^{-1}$.   

We thus highlight some channels that will be useful in establishing a $b'$ state, 
and decay channels that reveal its vector-like nature.

\medskip
\noindent {\bfseries {\it\bfseries Acknowledgments}:} 
We thank G.~Moreau and R.~Singh for discussions on related work at the ``Physics at TeV Colliders'' Les Houches 2009 workshop, 
P.~Behera for discussions on $b$-tagging issues, and, K.~Agashe and T.~Han for comments on the manuscript.

\setcounter{section}{0}
\renewcommand\thesection{Appendix \Alph{section}}               
\renewcommand\thesubsection{\Alph{section}.\arabic{subsection}}
\renewcommand\thesubsubsection{\Alph{section}.\arabic{subsection}.\arabic{subsubsection}}

\renewcommand{\theequation}{\Alph{section}.\arabic{equation}}    
\renewcommand{\thetable}{\Alph{section}.\arabic{table}}          
\renewcommand{\thefigure}{\Alph{section}.\arabic{figure}}        


\end{document}